\documentclass[12pt]{article}
 
\usepackage{textcomp}
\usepackage[bottom=30mm, top=30mm, left=25mm,right=25mm]{geometry}
\usepackage{setspace}
\usepackage{cite}
\usepackage{amssymb,amsmath}
\usepackage{bbm}
\usepackage{tikz}
\usetikzlibrary{shapes.misc}
\tikzset{cross/.style={cross out, thick, draw=black, minimum size=20*(#1-\pgflinewidth), inner sep=0pt, outer sep=0pt}, cross/.default={1pt}}
\usepackage{graphicx}
\usepackage{xcolor}

\newcommand{\cL}{\mathcal{L}}
\newcommand{\bR}{\mathbb{R}}

\usepackage{hyperref}

\begin{document}
\numberwithin{equation}{section}
\setstretch{1.2}
~\vspace{1cm}~
\begin{center}
{\Large\textbf{Metric Flows with Neural Networks}}\\[10mm]

James Halverson$^{a,c,}$\footnote{\href{mailto:j.halverson@northeastern.edu}{j.halverson@northeastern.edu (corresponding author)}}, Fabian Ruehle$^{a,b,c,}$\footnote{\href{mailto:f.ruehle@northeastern.edu}{f.ruehle@northeastern.edu}}\\[10mm]
{
	{\it ${}^{\text{a}}$ Department of Physics, Northeastern University, Boston, MA 02115, USA}\\[.5em]
	{\it ${}^{\text{b}}$ Department of Mathematics, Northeastern University, Boston, MA 02115, USA}\\[.5em]
	{\it ${}^{\text{c}}$ NSF Institute for Artificial Intelligence and Fundamental Interactions}\\[.5em]
}
\end{center}
\setcounter{footnote}{0}
\vspace{24pt}
\begin{abstract}
We develop a general theory of flows in the space of Riemannian metrics induced by neural network gradient descent. This is motivated in part by recent advances in approximating Calabi-Yau metrics with neural networks and is enabled by recent advances in understanding flows in the space of neural networks. We derive the corresponding metric flow equations, which are governed by a metric neural tangent kernel, a complicated, non-local object that evolves in time. However, many  architectures admit an infinite-width limit in which the kernel becomes fixed and the dynamics simplify. Additional assumptions can induce locality in the flow, which allows for the realization of Perelman's formulation of Ricci flow that was used to resolve the 3d Poincar\'e conjecture. We demonstrate that such fixed  kernel regimes lead to poor learning of numerical Calabi-Yau metrics, as is expected since the associated neural networks do not learn features.  Conversely, we demonstrate  that well-learned numerical metrics at finite-width exhibit an evolving metric-NTK, associated with feature learning. Our theory of neural network metric flows therefore explains why neural networks are better at learning Calabi-Yau metrics than fixed kernel methods, such as the Ricci flow.
\end{abstract}
\clearpage
\tableofcontents
\clearpage

\section{Introduction}
There are no known nontrivial compact Calabi-Yau metrics, objects of central importance in string theory and algebraic geometry, despite decades of study.

The essence of the problem is that theorems by Calabi~\cite{Calabi:1957aaa} and Yau~\cite{Yau:1977aaa,Yau:1978aaa} guarantee the existence of a Ricci-flat K\"ahler metric (Calabi-Yau metric) when certain criteria are satisfied, but Yau's proof is non-constructive. It is not for lack of examples satisfying the criteria, since topological constructions ensure the existence of an exponentially large number of examples~\cite{Candelas:1987kf,Kreuzer:2000xy,Halverson:2017ffz}. The problem also does not prevent certain types of progress in string theory, since aspects of Calabi-Yau manifolds can be studied without knowing the metric. For instance, much is known about volumes of calibrated submanifolds~\cite{HarveyLawson}, an artifact of supersymmetry and the existence of BPS objects, as well as metric deformations that preserve Ricci-flatness, the (in)famous moduli spaces~\cite{Candelas:1989bb}. Nevertheless, the central problem persists: general geometric properties of manifolds, and associated physics arising from compactification, requires knowing the metric. 

For this reason, it is interesting to study approximations of Calabi-Yau metrics. Efforts to do so generally require a sequence of approximations that converge toward the desired metric, which can be thought of as a flow in the space of metrics if the sequence is continuous. A classical example with a discrete sequence is Donaldson's algorithm, which uses a balanced metric to converge to the Calabi-Yau metric \cite{Donaldson:2005aaa}. More recently, there has been progress in approximating Calabi-Yau metrics with neural networks ~\cite{Anderson:2020hux,Douglas:2020hpv,Jejjala:2020wcc,Larfors:2021pbb,Larfors:2022nep,Gerdes:2022nzr}, which is the current state-of-the-art and is a continuous metric flow in the limit of infinitesimal update step size. Thirty minutes on a modern laptop gives an approximate Calabi-Yau metric on par with those that would take a decade to compute with Donaldson's algorithm.

The success of these neural network techniques prompts a number of mathematical questions. What is the mathematical theory underlying flows in the space of metrics induced by continuous neural network gradient descent? Since empirical  results suggest that the flows are converging to the Calabi-Yau metric, how does the flow relate to other flows that have Calabi-Yau metrics as fixed points, such as the Ricci flow?

The main result of this paper is to develop a theory of metric flows induced by neural network gradient descent, utilizing a recent result from ML theory known as the neural tangent kernel (NTK)~\cite{Jacot:2018aaa,Lee:2019aaa}. We will derive the relevant flow equations and demonstrate that a metric-NTK governs the flow, which in general is non-local and evolves in time. However, many architectures~\cite{yang2020tensor} admit a certain large-parameter limit~\cite{Jacot:2018aaa} in which the dynamics simplify and the metric-NTK becomes constant. Additional choices may be made to induce locality in the flow, and an appropriate choice of loss function reproduces Perelman's formulation of Ricci flow that was utilized to resolve the 3D Poincar\'e conjecture.

However, we will see that the assumptions that realize Ricci flow seem both ad hoc and strong from a neural network perspective, suggesting that it is very non-generic in the space of neural network metric flows. We will situate it within the general theory we develop and demonstrate experimentally that related fixed kernel methods lead to suboptimal CY metric learning. Thus, neural network metric flows provide a rich generalization of certain flows in the differential geometry literature, and the more general neural networks --- such as those of recent empirical successes --- make weaker assumptions and outperform fixed kernel methods. 

We will elaborate on this interplay extensively in the Conclusion.

\section{Metric Flows with Neural Networks} 
In this section we will develop a theory of flows in the space of metrics under the assumption that it is represented by a neural network $g_\theta$, where $\theta$ are the parameters of the neural network and a flow in $g_\theta$ is induced by a flow in $\theta$. 

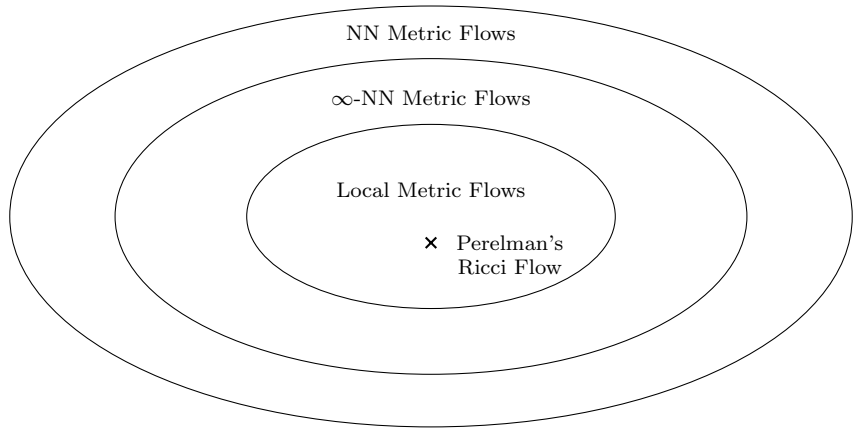
\begin{figure}
    \centering
\begin{tikzpicture}[scale=.7]
    \draw[] (0,0) ellipse (8cm and 4cm);
    \draw[] (0,0) ellipse (5.8cm and 2.8cm);
    \draw[] (0,0) ellipse (3.5cm and 1.75cm);
    \node[] at (0,3.5cm) {\scriptsize NN Metric Flows};
    \node[] at (0,2.25cm) {\scriptsize $\infty$-NN Metric Flows};
    \node[] at (0,.5cm) {\scriptsize Local Metric Flows};
    \draw (0,-.5cm) node[cross,black] {};
    \node[] at (1.5cm,-.5cm) {\scriptsize Perelman's};
    \node[] at (1.5cm,-.96cm) {\scriptsize Ricci Flow};
    \node[] at (-6.8cm,-.5cm) {\scriptsize Finite NN};
    \node[] at (-6.6cm,-.96cm) {\scriptsize CY Metric};
    \node[] at (-6.4cm,-1.42cm) {\scriptsize Successes};
\end{tikzpicture}
\caption{Graphical depiction of the different types of NN metric flows developed in this paper. Perelman's formulation of Ricci flow is realized as an infinite neural network metric flow under architectural assumptions that induce locality. Recent empirical successes in learning Calabi-Yau metrics realize many different flows in the outer ring.}
\label{fig:flow summary}
\end{figure}

For clarity, we summarize the results of this section. We will focus on the case that the parameters $\theta$ are updated by gradient descent with respect to a scalar loss functional $\cL$, and derive the associated flow equations in a ``time'' parameter $t$. Without making further assumptions, the metric flow is non-local and is given by an integral differential equation that involves a $t$-dependent kernel. However, many neural network architectures have a hyperparameter $N$ such that the associated kernel becomes deterministic and $t$-independent in the limit  $N\to \infty$. This is a dramatic simplification of the dynamics, and we call such a metric flow an \emph{infinite neural network metric flow}; it is related to more traditional kernel methods, with the fixed kernel determined by the neural network architecture. An infinite NN metric flow is still non-local and exhibits a certain type of mixing, but under additional assumptions about the architecture locality is achieved and mixing is eliminated, which we call a \emph{local neural network metric flow}. Such a flow is a local gradient flow, which allows metric flows defined as gradient flows to be realized in a neural network context. In particular, we show that Perelman's formulation of Ricci flow as a gradient flow \cite{Perelman:2002aaa} is a local neural network metric flow. See Figure \ref{fig:flow summary} for a graphical summary of these various flows. We collect other types of gradient flows tailored towards K\"ahler metrics in Appendix~\ref{app:Flows}. Being gradient flows of scalar loss functionals, these flows are amenable to an analysis that parallels our discussion of Perelman's Ricci flow; indeed, Perelman's Ricci flow can be written in terms of energy functionals of K\"ahler-Ricci flow~\cite{Headrick:2009jz}, see Appendix~\ref{app:Flows}.

\subsection{General Theory and Metric-NTK \label{sec:NN-metric-flows}}
Consider a Riemannian manifold $X$ with metric $g$, given by $g_{ij}$ in some local coordinates. In this work we consider the case that $g_{ij}$ is represented by a deep neural network, i.e. it is a function composed of simpler functions, with a set of parameters $\theta_I$, where $I=1,\dots,N$; see Appendix \ref{app:NNPrimer}. Generally $N$ is large for modern neural networks, and in this work we will exploit a different description that emerges in the $N\to \infty$ limit. Throughout, we use lower case Latin indices for the coordinates of $X$, and capital Latin indices for the network parameters.
Written this way, the metric evolves as 
\begin{equation}
\frac{dg_{ij}(x)}{dt} = \frac{d g_{ij}(x)}{d\theta_I} \frac{d\theta_I}{dt},
\end{equation}
with Einstein summation implied for repeated indices, unless stated otherwise. Training the network means updating the parameters by some non-zero $d\theta_I/dt$, and therefore this equation says that the metric flow may be thought of as a flow induced by training the neural network. 

There are many mechanisms for training a neural network in the deep learning literature, but one of the most common is gradient descent, in which case the parameter update is
\begin{equation}
\frac{d\theta_I}{dt} = -\frac{\partial \mathcal{L}[g]}{\partial \theta_I},
\end{equation}
where $\cL[g]$ is a scalar loss functional that depends on the metric. 
The loss sometimes depends on a finite set of points $B$ on $X$ known as the batch, in which case 
\begin{equation}
\cL[g] = \sum_{x' \in B} l[g](x'),
\end{equation}
where $l[g](x')$ is a pointwise loss that still depends on the metric. Alternatively, the loss could be over all of $X$
\begin{equation}
\cL[g] = \int_X d\mu(x') \, l[g](x').
\end{equation}
where $d\mu(x')$ is a chosen measure on $X$. In typical machine learning applications the training data is fixed and drawn (implicitly) from a fixed data distribution, which in the continuous case corresponds to utilizing a fixed measure; in particular, below we will use a volume form with respect to a fixed reference metric $\tilde g$. Henceforth, we drop the $[g]$ on $\cL$ and $l(x')$; it is to be understood that they depend on $g$.

Either type of loss yields additional expressions for the metric flow induced by gradient descent. For the discrete and continuous case, we obtain
\begin{equation}
    \frac{dg_{ij}(x)}{dt} = - \frac{\partial g_{ij}(x)}{\partial\theta_I}\sum_{x' \in B}  \frac{\partial g_{kl}(x')}{\partial \theta_I}  \frac{\delta l(x')}{\delta g_{kl}(x')}
\end{equation}
and
\begin{equation}
    \frac{dg_{ij}(x)}{dt} = - \frac{\partial g_{ij}(x)}{\partial\theta_I} \int_X d\mu(x')\,   \frac{\partial g_{kl}(x')}{\partial \theta_I}  \frac{\delta l(x')}{\delta g_{kl}(x')}\,,
\end{equation}
respectively. These expressions are written in a way to emphasize that one of the $\partial g / \partial \theta$ factors is outside of the sum or integral. Pulling them inside the sum or integral, we see the appearance of a distinguished object, 
\begin{equation}
\label{eqn:metric-NTK}
\Theta_{ijkl}(x,x') := \frac{\partial g_{ij}(x)}{\partial\theta_I} \frac{\partial g_{kl}(x')}{\partial \theta_I},
\end{equation}
which is a type of Neural Tangent Kernel (NTK) that we call the metric-NTK, which is symmetric in the first and last pair of indices. In terms of the metric-NTK, the neural network metric flows are
\begin{align}
    \label{eqn:NTK_metric_flow}
    \begin{split}
    \frac{dg_{ij}(x)}{dt} &= - \sum_{x' \in B} \Theta_{ijkl}(x,x')  \frac{\delta l(x')}{\delta g_{kl}(x')} \\ 
    \frac{dg_{ij}(x)}{dt} &= - \int_X d\mu(x')\,  \Theta_{ijkl}(x,x')  \frac{\delta l(x')}{\delta g_{kl}(x')}
    \end{split}
\end{align}
for the discrete batch and continuous batch, respectively.  We emphasize that the metric-NTK $\Theta_{ijkl}$ is not a 4-tensor, since it depends on both $x$ and $x'$ : the $(ij)$ indices transform with respect to diffeomorphisms of the external point $x$, and the $(kl)$ indices transform with respect to diffeomorphisms of the integrated point $x'$, with invariance of metric updates under $x'$-diffeomorphisms requiring an invariant measure.

In general, this is a complicated metric flow. Some of the reasons include:
\begin{itemize}
    \item \textbf{Evolving Kernel.} The metric flow is induced by gradient descent, which updates the parameters and therefore the metric-NTK; $\Theta$ and $\delta l / \delta g$ in \eqref{eqn:NTK_metric_flow} are time-dependent. 
    \item \textbf{Non-locality.} In general, $\Theta$ induces non-local dynamics: it is a smearing function that updates the metric at $x$ according to properties at other points $y$, which in principle could be far away from $x$.
    \item \textbf{Component Mixing.} We see that any $(k,l)$ components of the metric may update fixed $(i,j)$ components of the metric via mixing of $\Theta_{ijkl}$ and $\delta l/\delta g_{kl}$.
\end{itemize}
In the continuous case we have an integral differential equation that is difficult to solve and analyze. We will identify cases in which the situation simplifies. 

\subsection{Infinite Neural Network Metric Flows \label{subsec:inf-NN-metric-flows}}

In certain limits the metric-NTK enjoys properties that simplify the neural network metric flow. We utilize the two central observations from the NTK literature~\cite{Jacot:2018aaa,Lee:2019aaa} and refer the reader to Section \ref{eqn:concrete_arch} for details in an example; we review only the essentials here.

First, for appropriately chosen architecture (including normalization), the metric-NTK in the $N \to \infty$ limit may be interpreted as an expectation value by the law of large numbers, in which case the associated integral over parameters renders it parameter-independent. Schematically, we have
\begin{equation}
    \lim_{N\to \infty} \Theta_{ijkl}(x,x') = \mathbb{E}_\theta[\alpha_{ijkl}(x,x')] =: \bar \Theta_{ijkl}(x,x'),
\end{equation}
where the bar over $\bar \Theta$ reminds us that the metric-NTK in this limit is an average (over parameters) of a tensor $\alpha_{ijkl}(x,x')$ that may be computed in examples. While $\Theta$ is stochastic, due to its dependence on the parameters associated to some initial neural network draw, $\bar \Theta$ is deterministic, depending only on the network architecture and parameter distribution. 

The second simplification occurs for linearized models (linearized in the parameters, not the input $x$), defined to be 
\begin{equation}
g^L_{ij}(x) := g_{ij}(x)\Big|_{\theta = \theta_0} + \left(\theta_I - \theta_{0,I} \right) \frac{\partial g_{ij}(x)}{\partial \theta_I}\Big|_{\theta = \theta_0},
\end{equation}
i.e., it is just the Taylor expansion in parameters, truncated at linear order. The metric-NTK associated to the linear model is 
\begin{equation}
    \Theta^L_{ijkl}(x,x') = \frac{\partial g_{ij}(x)}{\partial \theta_I}\Big|_{\theta = \theta_0} \frac{\partial g_{kl}(x')}{\partial \theta_I}\Big|_{\theta = \theta_0} = \Theta_{ijkl}(x,x')\Big|_{\theta=\theta_0}.
\end{equation}
It is the metric-NTK $\Theta$ associated to $g$, evaluated at initialization. That is, though $\Theta$ evolves in $t$, $\Theta^L$ does not. Taking the $N\to \infty$ limit of the linearized model, its metric-NTK $\bar \Theta^L$ is both $t$-independent and $\theta$-independent.  $\bar \Theta^L$ is the so-called ``frozen" NTK.

Though linearization may seem like a violent truncation, it has been shown that deep neural networks evolve as linear models \cite{Lee:2019aaa} in the $N\to \infty$ limit, i.e.,  $\bar \Theta^L$ governs its gradient descent dynamics to a controllable approximation. This regime is known as ``lazy learning.'' For this reason, we henceforth drop the superscript $L$ and write the frozen NTK as $\bar \Theta$. In general, any quantity with a bar is frozen, i.e., deterministic and $t$-independent.

In summary, in the frozen-NTK limit the general NN metric flow \eqref{eqn:NTK_metric_flow} becomes
\begin{align}
    \label{eqn:inf_NTK_metric_flow}
    \begin{split}
    \frac{dg_{ij}(x)}{dt} &= - \sum_{x' \in B} \bar \Theta_{ijkl}(x,x')  \frac{\delta l(x')}{\delta g_{kl}(x')}\,, \\ 
    \frac{dg_{ij}(x)}{dt} &= - \int_X d\mu(x')\,  \bar \Theta_{ijkl}(x,x')  \frac{\delta l(x')}{\delta g_{kl}(x')}\,,
    \end{split}
\end{align}
in the discrete and continuous case,
with the only difference being that the dynamics are governed by the deterministic $t$-independent kernel $\bar \Theta$ rather than the generally stochastic $t$-dependent kernel $\Theta$. The dynamics still exhibits non-locality and component mixing, but no longer has an evolving kernel. We refer to such a flow as an \emph{infinite neural network metric flow}.

Though the equations have only changed by the introduction of the bar, this is actually a dramatic simplification of the dynamics! Naively, one might think that although the kernel is better behaved, one cannot train an infinite neural network because one cannot put it on a computer. This is in fact not true: while initializing an infinite NN and running parameter space gradient descent is indeed impossible, the large-$N$ limit gives a new description of the system --- a different duality frame --- that allows to analyze the same dynamics in a different way that does not require explicitly updating an infinite number of a parameters. For instance, for mean-squared-error loss the dynamics may be solved \emph{exactly}, with computable mean and covariance across different initializations~\cite{Lee:2019aaa}. This is the sort of behavior expected of duality: by describing the same system in a different way, here as kernel regression on function space rather than neural network gradient descent in parameter space, new calculations become possible that would be impossible in the original description. In the mentioned MSE case, the calculations amount to the average prediction and covariance of an infinite number of infinitely wide neural networks trained to infinite time; clearly this is infeasible in the infinite dimensional parameter space description!

\subsection{Local Metric Flows\label{subsec:local metric flows}}
Let us discuss how to simplify the flows even further by getting rid of non-locality and component mixing. While such flows are less general, they are of interest because they are more analytically tractable, and we will also see that they recover some famous metric flows.

To simplify matters, we will consider local metric flows that do not exhibit component mixing, though the latter could be relaxed. Specifically, if the discrete and continuous frozen metric-NTK satisfy
\begin{align}
\label{eqn:local_metric_NTK}
\begin{split}
\bar \Theta_{ijkl}(x,x') &= \delta_{x,x'} \,\,\delta_{ik} \delta_{jl}\,\,\bar \Omega(x)\,, \\ 
\bar \Theta_{ijkl}(x,x') &= \delta(x-x') \,\, \delta_{ik} \delta_{jl}\,\,\bar \Omega(x)\,,
\end{split}
\end{align}
respectively, for some deterministic function $\bar \Omega(x)$, the dynamics becomes
\begin{align}
    \label{eqn:local_metric_flow}
    \frac{dg_{ij}(x)}{dt} = - \bar \Omega(x)  \frac{\delta l(x)}{\delta g_{ij}(x)}\,.
\end{align}
Here $\delta_{x,x'}$ is the discrete Kronecker delta and $\delta(x-x')$ is the Dirac delta function. We call such a flow a \emph{local metric flow}.

However, there is a pathology that must be discussed.  In deep learning, one typically splits the input into disjoint sets containing train inputs $x'$ and test inputs $x$, respectively. NN updates are trained on train inputs $x'$ only and their generalization properties to unseen points is tested with the test set inputs $x$. In the continuous case, there simply is no disjoint test set since we integrate over the entire manifold $X$. In the discrete case, enforcing delta-functions in the frozen metric NTK in~\eqref{eqn:local_metric_NTK} means that only train points are updated and there would be no learning on any point outside the train set. Hence, a non-local metric-NTK is essential to ensuring that the metric gets updated at all manifold points. Thus, the local case~\eqref{eqn:local_metric_flow} only makes sense for continuous flows.

Finally, there is a simple trick for defining a new architecture that gets rid of $\bar\Omega(x)$ or reshapes it, if desired. Such a choice would remove the $\bar\Omega(x)$ dependence of the architecture, but the locality and component mixing choices of the architecture remain intact. We will phrase the trick in terms of a general neural network, but apply it in the context of metrics.

Let $\phi(x)$ be any network function with associated NTK $\Theta_\phi(x,x')$. Now multiply $\phi(x)$ by a \emph{deterministic} function $h(x)$ to obtain a new network 
\begin{equation}
    \widetilde\phi(x) = h(x)\phi(x).
\end{equation}
Since $h(x)$ is deterministic, $\phi$ and $\widetilde\phi$ have the same parameters and
\begin{equation}
\Theta_{\widetilde\phi}(x,x') = h(x)h(x') \Theta_{\phi}(x,x').
\end{equation}
This also gives the same result for frozen NTKs,
$\bar \Theta_{\widetilde\phi}(x,x') = h(x)h(x') \bar \Theta_{\widetilde\phi}(x,x')$. In the case of a local flow,
where we have 
\begin{equation}
\Theta_{\phi}(x,x') = \delta(x-x') \,\Omega(x), 
\end{equation}
the $\delta$-function imposes that the two $h$'s are evaluated at the same point, yielding 
\begin{equation}
    \Theta_{\tilde \phi}(x,x') = \delta(x-x')\, h^2(x) \,\Omega(x). 
\end{equation}
This trick is useful because we may use the freedom of $h(x)$ to define a new architecture $\widetilde\phi$ whose NTK we can shape, and in particular cancel out potentially unwanted local factors such as $\Omega(x)$ in the NTK or $\sqrt{g(x')}$ in the volume form.

\subsubsection{Architecture Design for Local Metric Flow}

We have seen that a metric-NTK satisfying \eqref{eqn:local_metric_NTK} induces a local metric flow \eqref{eqn:local_metric_flow} that evolves with a deterministic kernel, a local evolution equation, and without mixing induced by $\Theta_{ijkl}$ and $\delta l/\delta g_{kl}$. In this section, we study whether NN architectures that satisfy \eqref{eqn:local_metric_NTK} actually exist. Specifically, we investigate how the different delta functions may arise in the metric-NTK.

There is a simple sufficient condition for obtaining the Kronecker deltas $\delta_{ik} \delta_{jl}$.\footnote{These Kronecker deltas amount to a choice of gauge that is not preserved under diffeomorphisms in $x$ and $x'$ on the metric-NTK.} Choosing each independent component of the metric $g_{ij}(x)$ to be a separate neural network with parameters $\theta^{ij}$, the 
set of parameters $\theta$ for the entire metric is partitioned as 
\begin{equation}
    \theta = \bigcup_{i,j} \theta^{ij},
\end{equation}
and the metric-NTK is
\begin{equation}
\bar \Theta_{ijkl}(x,x') = \sum_{m,n} \frac{\partial g_{ij}(x)}{\partial \theta^{mn}_I}\frac{\partial g_{kl}(x')}{\partial \theta^{mn}_I} = \delta_{ik}\delta_{jl} \, \frac{\partial g_{ij}(x)}{\partial \theta^{ij}_I}\frac{\partial g_{kl}(x')}{\partial \theta^{ij}_I},
\end{equation}
with no Einstein summation over lower-case Latin indices on the RHS.
The last pair of derivatives is the NTK $\bar \Theta^{(ij)}(x,x')$ of the $(i,j)$ component of the metric, giving
\begin{equation}
    \label{eqn:independent component NTK}
    \bar \Theta_{ijkl}(x,x') = \delta_{ik}\delta_{jl} \, \bar \Theta^{ij}(x,x'),
\end{equation}
Thus, the metric-NTK for a metric with components given by independent neural networks evolves according to the NTK of the individual components, as expected.
By symmetry, we should choose the independent neural networks associated to each component to have identical architecture. Since the architectures for the components are the same, they have the same frozen NTK,
\begin{equation}
\bar \Theta(x,x') := \bar \Theta^{ij}(x,x'),
\end{equation}
and we have that the frozen metric-NTK is 
\begin{equation}
    \label{eqn:NTK_no_mixings}
    \bar \Theta_{ijkl}(x,x') = \delta_{ik}\delta_{jl} \, \bar \Theta(x,x').
\end{equation}
This gets us part of the way to \eqref{eqn:local_metric_NTK}; we have the Kronecker deltas that prevent component-mixing.

We still need the Dirac delta function that induces locality, however. This is a non-trivial step. Given \eqref{eqn:NTK_no_mixings}, we must find an architecture such that 
\begin{equation}
    \label{eqn:X_deltafunc_omegabar}
\bar \Theta(x,x') = \delta(x-x')\, \bar \Omega(x)
\end{equation}
for some $\bar \Omega(x)$.
The $\delta$-function is defined with respect to the measure $d\mu(x')$, which for simplicity we
take to be the volume form with respect to a fixed reference metric $\tilde g$ on $X$, although the same argument applies for a more general probability density $d\mu(x') = d^d x'\, P(x')$. Since the $\delta$-function has support on a local patch $\mathbb{R}^d$, we have 
\begin{equation}
    \label{eqn:delta on X}
\int_X dV_{\tilde g} \,\delta(x-x') f(x') =
\int_{\mathbb{R}^d} d^d x' \sqrt{|\tilde g(x')|}\, \delta(x-x')\, f(x') = f(x).
\end{equation}
This $\delta$-function on $X$ is related to the $\delta$-function on $\mathbb{R}^d$ by 
\begin{equation}
\delta_{\mathbb{R}^d}(x-x') = \sqrt{|\tilde g(x')|}\, \delta(x-x'),
\end{equation}
which gives the usual identity $\int d^dx' \delta_{\mathbb{R}^d}(x-x')f(x')=f(x)$ when inserted into~\eqref{eqn:delta on X}.

We take a two-step process to obtain \eqref{eqn:X_deltafunc_omegabar}: we must figure out how to obtain $\delta_{\bR^d}$ inside an NTK, and then how to account for the factor $\sqrt{|g|}$ in the volume measure. Let $\phi_\sigma(x)$ with $\sigma\in\mathbbm{R}^+$ be a network with frozen NTK $\bar \Theta_\sigma(x,x')$ given by
\begin{equation}
\bar \Theta_\sigma(x,x') = \frac{1}{(2\pi \sigma^2)^{d/2}} \exp^{-\frac12 \frac{|x-x'|^2}{\sigma^2}} \,\, \bar \alpha(x,x')
\label{eqn:gaussianalphantk}
\end{equation}
for some deterministic function $\bar \alpha(x,x')$. We emphasize that the semi-locality induced by the Gaussian suppression is \emph{not} general, and is another assumption that must be made to push towards local flows and (eventually) Perelman's formulation of Ricci flow. Clearly, we get the $\delta_{\bR^d}$ inside the NTK for $\sigma \to 0$.
Using the rescaling trick introduced in the previous section, we may define a new network function that is $\phi_\sigma(x)$ multiplied by $\tilde g(x)^{-1/4}$. This gives a new NTK that by abuse of notation we again call $\bar \Theta_\sigma$, given by 
\begin{equation}
    \label{eqn:parameterically_nonlocal_NTK}
    \bar \Theta_\sigma(x,x') = \frac{1}{(2\pi \sigma^2)^{d/2}} (\tilde g(x)\tilde g(x'))^{-1/4} \exp^{-\frac12 \frac{|x-x'|^2}{\sigma^2}} \,\, \bar \alpha(x,x'),
\end{equation}
where the only difference is the $\tilde g$-factors, by design.
This new NTK satisfies
\begin{equation}
\lim_{\sigma \to 0} \,\bar \Theta_\sigma(x,x') = \delta(x-x') \,\,\bar \Omega(x),
\end{equation}
i.e., it satisfies \eqref{eqn:X_deltafunc_omegabar} with $\bar \Omega(x) = \bar \alpha(x,x)$, where this $\delta$-function is with respect to the non-trivial volume measure. The parameter $\sigma \in \bR^+$ clearly sets a scale of non-locality in the network evolution, which is an interesting object to study. The family has an NTK given by \eqref{eqn:parameterically_nonlocal_NTK}, which induces a metric flow
\begin{equation}
    \label{eqn:parameterically non-local flow dynamics}
    \frac{dg_{ij}(x)}{dt} = - \int \, d^d x \left(\frac{\tilde g(x')}{\tilde g(x)}\right)^{1/4}\,\, \frac{1}{(2\pi \sigma^2)^{d/2}} \exp^{-\frac12 \frac{|x-x'|^2}{\sigma^2}} \bar\alpha(x,x')  \,\, \frac{\delta l(x')}{\delta g_{ij}(x')}.
\end{equation}
Here we see that the variance of the Gaussian $\sigma$ is a parameter controlling the amount of non-locality in the dynamics, since it affects how strongly updates at $x'$ affects the evolution of the metric at $x$. In the limit $\sigma \to 0$, the normalized Gaussian becomes the $\delta$-function and the dynamics is 
\begin{equation}
    \frac{dg_{ij}(x)}{dt} = -  \bar \Omega(x)  \,\, \frac{\delta l(x')}{\delta g_{ij}(x')},
\end{equation}
where $\bar \Omega(x)=\bar\alpha(x,x)$, and we have recovered a local metric flow. 

Summarizing, we obtain a network with the desired properties as follows:
\begin{itemize}
\item \textbf{No component mixing} between $\Theta_{ijkl}$ and $\delta l / \delta g_{kl}$ is obtained by taking each component to be a neural network of the same architecture, but with independent parameters.
\item \textbf{Locality} arises from a $\delta$-function as in \eqref{eqn:X_deltafunc_omegabar}, which we achieve by first obtaining a $\delta_{\bR^d}$ in the NTK as the limit of a normalized Gaussian, and then passing to the $\delta$-function on the Riemannian manifold by using the rescaling trick to obtain the geometric factor in the volume form.
\end{itemize}

\subsubsection{Concrete Architectures for Local and Non-local Metric Flows \label{eqn:concrete_arch}}
We have just shown that any family of architectures with parameter $\sigma$ satisfying \eqref{eqn:gaussianalphantk} may be rescaled to give a local metric flow as $\sigma \to 0$. Since we are considering cases where each metric component is independent, it suffices to consider scalar network functions. 

Consider a network architecture based on cosine activation, with 
\begin{align}
\begin{split}
\phi(x) = \frac{A}{\sqrt{N}} \sum_{i=1}^N \sum_{j=1}^d a_i \cos(w_{ij}x_j + b_i), \\  
a\sim \mathcal{N}(0,\sigma_a^2)\,,\qquad  w\sim \mathcal{N}(0,\sigma_w^2/d)\,, \qquad b\sim \mathcal{U}(-\pi,\pi)\,,
\end{split}
\end{align}
where $N$ is the width of the network and $A$ is a normalization factor that will be fixed later. To simplify the picture, we freeze all weights to their initialization values except the $a_i$, in which case the NTK is 
\begin{equation}
\Theta(x,x') = \sum_{k=1}^N \frac{\partial \phi(x)}{\partial a_k}\frac{\partial \phi(x')}{\partial a_k} = \frac{A^2}{N} \sum_{k=1}^N \sum_{j=1}^d \cos(w_{kj}x_j + b_k)\cos(w_{kj}x'_j + b_k)\,.
\end{equation}
As we take the width $N\to \infty$, the NTK $\Theta$ becomes an expectation value, by the law of large numbers,
\begin{equation}
\bar \Theta(x,x') = A^2\,  \mathbb{E}[\cos(w_{kj}x_j + b_k)\cos(w_{kj}x'_j + b_k)].
\end{equation}
Evaluating the expectation gives 
\begin{equation}
    \bar \Theta(x,x') = \frac{A^2}{2} \exp\left(-\frac12 \frac{\sigma_w^2}{d} |x-x'|^2\right).
\end{equation}
We see that our NTK is a Gaussian, as desired. Normalizing it, so that in an appropriate limit it is a $\delta$-function, we have (with $\sigma_w^2 = d \,\mathbb{E}[w_{ij}^2]$)
\begin{equation}
A = 2^\frac14 \left(\frac{\sigma_w}{\sqrt{2\pi d}}\right)^{d/4}.
\end{equation}
With this normalization factor fixed, the NTK is a normalized Gaussian of width
\begin{equation}
\sigma = \frac{\sqrt{d}}{\sigma_w}.
\end{equation}
We obtain a local metric flow by taking the limit $\sigma_w \to \infty$,
which sends $\sigma \to 0$.

The trick of freezing all weights but those of the last (linear) layer is common in the literature. In such a case the NTK is determined (up to a constant) by the so-called NNGP kernel $\mathbb{E}[\phi(x)\phi(x')]$, also known as the two-point correlation function. The architecture that we have presented is a simple modification (via so-called NTK parameterization, which gives the $1/\sqrt{N}$ factor) of an architecture in \cite{Halverson:2021aot} that is known to have a Gaussian NNGP kernel. Similarly, the architecture called Gauss-net in~\cite{Halverson:2020trp} has a Gaussian NNGP kernel, and can be appropriately modified to give a different architecture (based on exponentials in the network function, rather than cosines) that yields a local metric flow.

\subsection{Ricci Flow and More General Flows as Neural Network Metric Flows}

Having developed a theory of neural network metric flows, we ask: are any canonical metric flows from differential geometry realized as neural network metric flows? 

Many well-studied metric flows are local: they are not an integral PDE and take the form 
\begin{equation}
\label{eqn:u_metric_flow}
\frac{dg_{ij}(x)}{dt} = -u_{ij}(x),
\end{equation}
with metric updates depending on the local properties of some rank two symmetric tensor $u_{ij}(x)$. A particularly well-studied example is Ricci flow
\begin{equation}
    \label{eqn:Ricci flow}
    \frac{dg_{ij}(x)}{dt} = -2R_{ij}(x).
\end{equation}
In general, flows of the form \eqref{eqn:u_metric_flow} are not necessarily gradient flows. To make a direct connection, one could study neural network metric flows that are not gradient flows either, but following a standard assumption in deep learning  we studied those neural network metric flows that are induced by the gradient of a scalar loss functional.

Famously, Perelman showed~\cite{Perelman:2002aaa} that Ricci flow is a gradient flow after applying a $t$-dependent diffeomorphism (see~\cite{Kleiner:2008aaa} for a review).
Specifically,
\begin{equation}
    \label{eqn:Ricci flow with f}
    \frac{dg_{ij}}{dt} = \frac{\delta \mathcal{F}[\phi,g]}{\delta g_{ij}(x)} = -2[R_{ij} + \nabla_i \nabla_j \phi]\,,
\end{equation}
where 
\begin{equation}
\label{eqn:PerelmanFunctional}
\mathcal{F}[\phi,g] = \int_X \left( R + |\nabla \phi|^2 \right)\, e^{-\phi}\, dV
\end{equation}
is the Perelman functional and \eqref{eqn:Ricci flow with f} is equivalent to \eqref{eqn:Ricci flow} by a $t$-dependent diffeomorphism, and $\phi$ is a scalar function that is known as the dilaton in string theory. Using our formulation, local metric flow induced by neural network gradient descent, e.g.\ via the concrete architecture of Section \eqref{eqn:concrete_arch}, we obtain Perelman's formulation of Ricci flow by simply choosing the loss function appropriately,
\begin{align}
    l(x) = - \frac{\mathcal{F}[\phi,g]}{\bar \Omega(x)}\,.
\end{align}
Similarly, we obtain a parametrically non-local generalization of Ricci flow by backing off of the $\sigma \to 0$ limit. Doing so replaces the $\delta$-function in the NTK by a Gaussian, and allows nearby points $y$ to influence metric updates at $x$, in a way controllable by tuning $\sigma$.

Similarly, any other local metric flow that is a gradient flow may be achieved, together with its non-local generalization for finite $\sigma$, by the choice of loss function and a tuning of $\sigma$.

\section{Numerical Implementation}
In this section we describe the use of these (kernel) methods to approximate CY metrics numerically. We will use the standard test case of the Fermat Quintic to benchmark the methods. The Fermat Quintic is a Calabi-Yau threefold that is given as the anti-canonical hypersurface inside a complex projective space $\mathbbm{P}^4$ with homogeneous coordinates $z_i$ by the equation
\begin{align}
\label{eq:FermatQuintic}
z_0^5+z_1^5+z_2^5+z_3^5+z_4^5=0\,.
\end{align}
A simple NN with three layers, 64 nodes, and ReLU activation function can pick up a factor of 10-100 in the overall validation loss, when trained for around 50 epochs on 100k points on the Fermat Quintic~\cite{Larfors:2021pbb,Larfors:2022nep}. This factor serves as a natural benchmark for comparison with our kernel methods.

The main takeaway is that kernel methods work excellently on the train set, but generalization to test points is challenging. One reason is that there are multiple numerical challenges associated with the implementation. We will describe how we overcome these by implementing feature clustering. We then evaluate performance using ReLU, Gaussian, and semi-local kernels. In all cases, test set performance lacks behind that of simple (finite width) NNs. We attribute this lack of performance of frozen metric NTK methods to the importance of feature learning, which is impossible for frozen NTKs. Indeed, feature learning necessitates that the kernel corresponding to the NN adapts to features, i.e., evolves during training. 

\subsection{Clustering and Preprocessing}
We study the discrete NTK $\Theta_{ijkl}(x_a,x'_b)$, where $i,j,k,l\in\{1,2,3\}$. The index $b$ always runs over the $N_\text{pts}'$ train points, $b\in\{1,\ldots,N_\text{pts}'\}$, whereas $a$ runs over the same set during clustering and training, but over labeling the $N_\text{pts}$ test point during prediction / inference with $a\in\{1,\ldots,N_\text{pts}\}$. During training, when we use all $N_\text{pts}'$ points, this is a $3\times3\times3\times3\times N_\text{pts}'\times N_\text{pts}'$ tensor, which has almost a trillion entries for $N_\text{pts}'=100,000$. This is a problem generic to NTK implementations~\cite{novak2019neural,lee2020finite,novak2022fast} and means that hardware far beyond academic grade is required for the full computation.

Instead of computing the full trillion-parameter tensor, we cluster the input features. For the problem at hand, it might be reasonable to assume that the metric at some test point $x'$ is most strongly correlated with the metric at nearby points $x$. Typically, there is no canonical metric on feature space, so the notion of distance is somewhat arbitrary. In our case, however, the features are actual points on a CY manifold, so a good notion of distance would be the shortest geodesic distance between these points with respect to the canonical (unique, Ricci-flat) metric. In practice, this is unfeasible for multiple reasons: First, we do not yet have the Ricci-flat CY metric -- we are computing it with the NTK. Second, even if we had the metric, computing the geodesic distance between two points requires solving an initial value PDE numerically (which can be done using e.g.\ shooting methods), but is very computationally costly. 

Hence, we resort to a much simpler proxy which is less accurate but cheap to compute: the shortest geodesic distance between two points in the ambient space Fubini-Study metric. The Fubini-Study metric is a canonical K\"ahler metric on complex projective spaces, see e.g.~\cite{Griffiths} for a mathematics introduction to the topic. For two points $z$ and $z'$ in $\mathbbm{P}^n$ given in terms of homogeneous coordinates, the geodesic distance with respect to the Fubini-Study metric is
\begin{align}
\label{eq}
d(z,z')=\arccos\sqrt\frac{(z^\dagger \cdot z')(z'^\dagger\cdot z)}{|z|^2 |z'|^2}\,,
\end{align}
where the dot product is with respect to the flat metric.

\begin{figure}
\centering
\includegraphics[width=.95\textwidth]{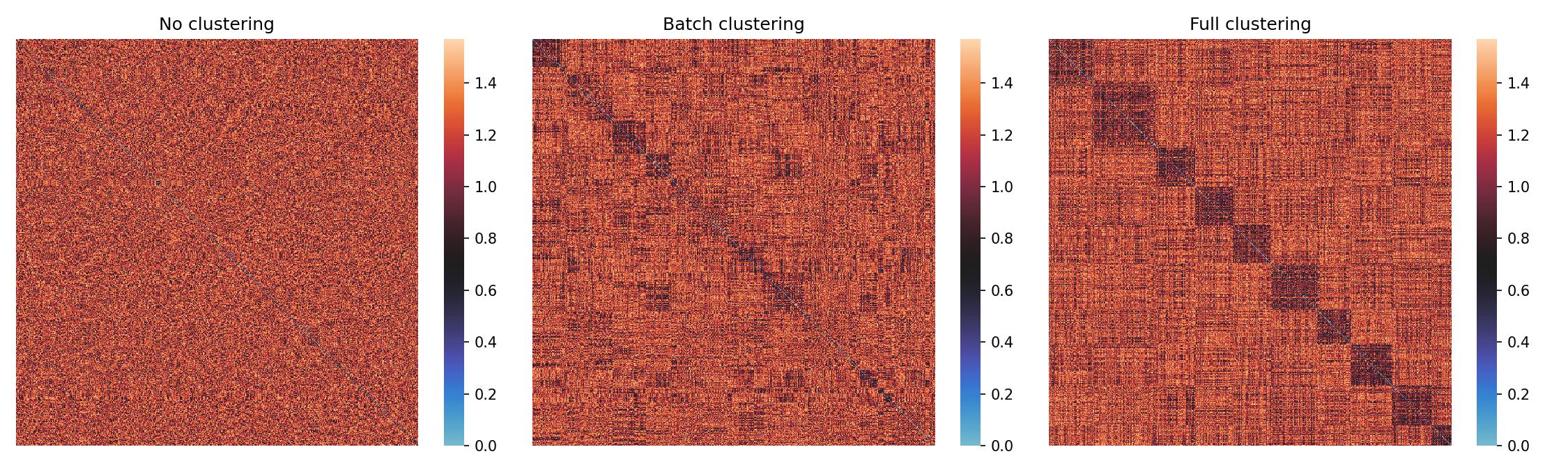}
\caption{Comparison of distance between points with no clustering, batch clustering and full clustering for 1000 points and 10 clusters, using 5 batches for batch clustering.}
\label{fig:HeatMap}
\end{figure}

In order to perform clustering based on distances for $N_\text{pts}'=100,000$ points, we would need to fit a $100,000\times100,000$ distance matrix into memory, which is also unfeasible. Instead, we implement the following procedure:
\begin{enumerate}
\item Divide the points into mini batches. The size of these should be adapted to the available hardware. We used a size of $M=10,000$.
\item Cluster each mini batch independently based on the FS geodesic distance. We use agglomerative clustering and average to compute the linkage of clusters, as implemented by \texttt{sklearn} \cite{scikit-learn}. Again the total cluster size depends on the available hardware. We aimed for $\mathcal{O}(5000)$ points in each cluster, which allowed us to compute the $3\times3\times3\times3\times5000\times5000$ NTK tensor, so we aimed for $C=N_\text{pts}'/5000$ clusters.
\end{enumerate}
At this point, we have $B=N_\text{pts}'/M$ batches (which we label by $I=1,2,\ldots B$), each of which has $C$ clusters $c^I_\alpha$, $\alpha=1,2,\ldots C$. Next, we need to merge the clusters across the different mini batches. To do so, we proceed as follows:
\begin{enumerate}
\setcounter{enumi}{2}
\item Use the clusters of the first mini batch as a starting point
\item For each cluster in mini batch 2, compute the geodesic distance between all points in this mini batch and the first cluster. The cluster from mini batch 2, whose mean distance over all points in cluster 1 of mini batch 1 is smallest, gets merged into cluster 1 of mini batch 1. 
\item Continue for all clusters in all mini batches.
\end{enumerate}
This means that the final clusters $c_\alpha$ are given by
\begin{align}
\label{eq:Clusters}
c_\alpha = \bigcup_\text{mini batches $I$}\min\left[\text{mean}_{\text{clusters $\beta$}}(d_\text{FS}(c_\alpha^1,c_\beta^I))\right]\,.
\end{align}
Based on the clusters, we can compute the NTK updates for each cluster separately, assuming that the contribution from other clusters can be neglected in the updates. While this algorithm is not perfect and depends on the random order of mini batches and clusters, it produces decent results as can be checked by comparing the result of this algorithm to the result when we just cluster without batching, which is feasible if the number of points is small enough. To illustrate the efficiency of the algorithm we present a heatmap of the distance matrix for 1000 points in Figure~\ref{fig:HeatMap}. On the left, we see the distance matrix without any clustering. In the middle, we give the distance matrix when ordering the points when clustered according to~\eqref{eq:Clusters} with 5 batches. On the right, we give the distance matrix when running the clustering algorithm on all 1000 points (i.e., without batching). To quantify the quality of the clustering algorithm we also computed the silhouette score~\cite{ROUSSEEUW198753} using \texttt{scikit-learn}~\cite{scikit-learn}. The silhouette score is a number between -1 and 1, with 1 indicating perfect clustering and values around 0 indicating overlapping clusters. We compute the silhouette score for the batched clustering with 5 batches, for the full clustering without batching, and for Birch clustering based on the Euclidean -- rather than Fubini-Study -- distance as a baseline (the result looks like no clustering to the naked eye). We obtain silhouette scores of .016, -.011, and -0.032 for the three cases, respectively.

There are further numerical subtleties that need to be taken into account. First, the input features are the real and imaginary part of the 5 homogeneous coordinates $z_i\in\mathbbm{P}^4$, $i=0,\ldots, 4$. In contrast, the CY metric is naturally expressed as a Hermitian $3\times3$ matrix, $g_{a\bar b}\;dx^a(z)\otimes dx^b(z)$, $a,b=1,2,3$. Here, the three complex coordinates $x_a$ are functions of the five homogeneous ambient space coordinates $z_i$. Typically, one chooses a coordinate system on the CY by going to affine coordinates in a patch of the ambient space (this removes one of the five $z_i$ coordinates through scaling) and eliminating one of the affine coordinates via the hypersurface equation that defines the CY in that patch. While physics does of course not depend on the choice of coordinate system, the metric $g_{a\bar b}$ will get modified by the Jacobian of the transformation from one coordinate system to another. 

Since computing the relevant data like the holomorphic $3$-form $\Omega$, the pullbacks, and the integration weights is prone to numerical errors, we use this freedom to choose patches and scalings that minimize the numerical error sources. In practice, this amounts to transforming to the patch where the coordinate $z_i$ with the largest absolute value is scaled to one, and then using the CY equation to eliminate the coordinate which results in the smallest value for $|\Omega|^2$. However, this means that there are 20 possible coordinate choices and the metric at different points will be in different coordinate systems. Hence, updating the matrix $g_{a\bar b}(y)$ using the matrix entries of $g_{i\bar j}(x)$ does not make any sense. One way around this issue is to transform the metric at each points into the same coordinate system. instead of doing this, we simply sample about 20 times more points than needed and then use rejection sampling to obtain a set of points whose metrics are in the same coordinate system under the above procedure. 

Another problem is a numerical instability related to the geodesic distance. Note that 
\begin{align}
\arccos(1-\epsilon) \simeq \sqrt{2\epsilon}
\end{align}
when $0<\epsilon\ll 1$.
This means that points $x$ and $y$ which are numerically identical (meaning $\epsilon\sim 10^{-7}$) would still have a sizable distance (.0003). Since our kernel exponentially suppresses updates at $x$ by $d(x,y)$, this leads to unwanted behavior during training. To ameliorate this, we set the geodesic distance to zero once it falls below a certain threshold, since we would only be amplifying numerical noise otherwise.

\subsection{Metrics from Kernels}

In this subsection we summarize a number of techniques that we used to approximate metrics on test points using kernels. We compute the metric updates for a discretized flow of~\eqref{eqn:NTK_metric_flow}, where we using first order backwards differentiation to compute the $(n+1)^\text{st}$ approximation for the metric using
\begin{align}
\label{eq:GDDiscrete}
g_{ij}^{(n+1)}(x)=g_{ij}^{(n)}(x) - \Delta t \sum_{x'\in B}\Theta_{ijkl}(x,x')\frac{\delta \ell(x')}
{\delta g_{kl}^{(n)}}\,,
\end{align}
where we treat the step size $\Delta t$ as a hyperparameter. For the NTK $\Theta_{ijkl}(x,x')$, we use different kernels, either computed from the \texttt{neural-tangents} package~\cite{novak2019neural} for a NN with ReLU activation function, or taken to be a fixed kernel (we study a Gaussian Kernel and a ``delta function'' or `nearest neighbor'' kernel) which are not necessarily induced by a (simple or obvious) NN architecture. For the loss $\ell(x')$, we use the sigma loss defined in~\eqref{eq:SigmaLoss}. To determine the hyperparameters (such as the learning rate $\Delta t$, the width of the Gaussian for the Gaussian kernel, etc.), we use a Python implementation~\cite{BayesianOptimization:2023aaa} of a Bayesian optimization scheme~\cite{Snoek:2012aaa}. We need to generate a new set of points for each Bayesian optimization run since we observed that otherwise the optimizer tunes the hyperparameters to the specific point set and shows poor performance on new data. For the optimal set of hyperparameters identified by the optimizer, we iterate~\eqref{eq:GDDiscrete} for 50 steps. After each iteration in~\eqref{eq:GDDiscrete}, we compute the sigma loss on the test set for the new metric $g_{ij}^{(n+1)}$.

\subsubsection*{Metrics from a ReLU NTK}

A simple single-layer network with ReLU activation may pickle up a factor of $10$ to $100$ in the total loss. We use this as a baseline of comparison for kernel methods, and specifically study the NTK associated to this ReLU network as computed by the python package \texttt{neural-tangents}~\cite{novak2019neural}, which has the ability to compute the infinite-width NTK of many architectures. 

\subsubsection*{Metrics from a Gaussian kernel}
We also use a simple Gaussian kernel, i.e., we set
\begin{align}
\Theta_{ijkl}(x,x')=\delta_{ik}\delta_{jl}e^{-\frac{d(x,x')}{2\sigma}}\,,
\end{align}
where the variance $\sigma$ is a hyperparameter and $d(x,x')$ is the shortest geodesic distance w.r.t.\ the ambient space FS metric. We recover the local update limit for $\sigma\to0$. In the finite $\sigma$ regime, the kernel updates at a point $x'$ are just the sums of all contributions from all training points $x$, weighted by a Gaussian proportional to their distance to $x'$. This suppression means that if the distance $d$ is larger than $\sigma$, the updates are tiny. For this reason, we include the possibility of normalizing the updates such that they are sizable at least at some points. 
Overall, we found that the optimizer prefers a narrow Gaussians. However, the performance is very sensitive to the hyperparameters and can easily lead to exploding gradients or metric updates that lead to non-positive definite matrices. When restricting the updates akin to gradient clipping, the performance suffers substantially.

\subsubsection*{Metrics from a Delta kernel}
Given that the Bayesian optimizer prefers narrow Gaussians, we finally study a kernel which simply updates the metric at a test point $x'$ with the metric correction computed for the closest point $x$ in the training set. We call this the Delta kernel.

\subsubsection*{Results and Discussion}

With the techniques that we have described, we performed many experiments with different values of the various hyperparameters. Though we were able to obtain significant gains in the train loss, our experiments failed to achieve a gain of more than a factor of $\sim 5$ in the test loss. We would like to streamline the presentation of our results to help understand this failure to generalize to test points.

\begin{figure}[t]
\centering
\includegraphics[width=.48 \textwidth]{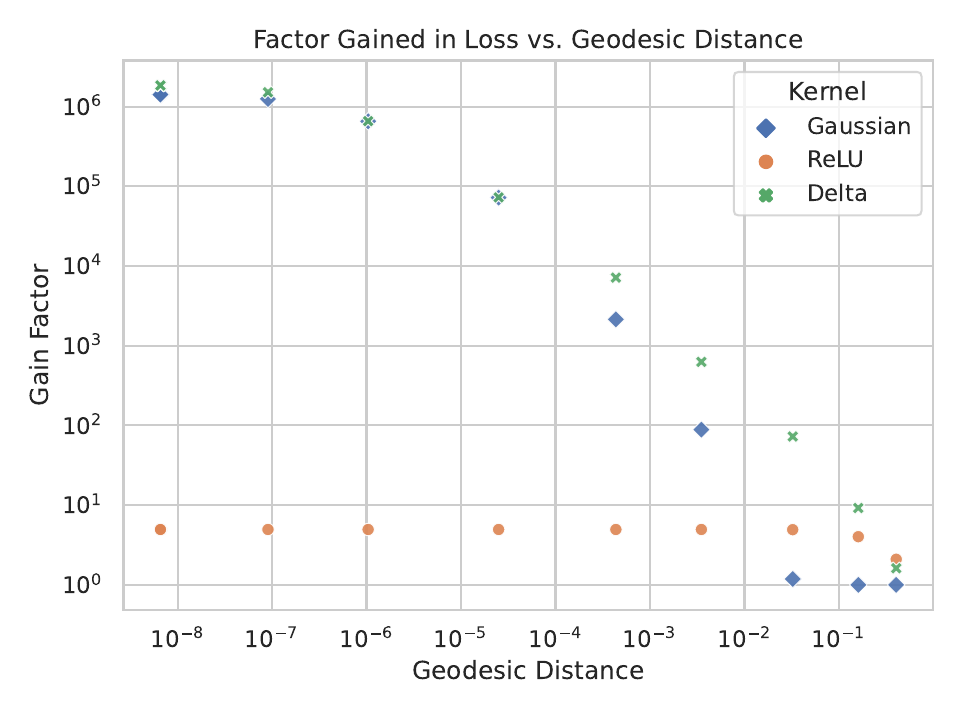} \quad 
\includegraphics[width=.48 \textwidth]{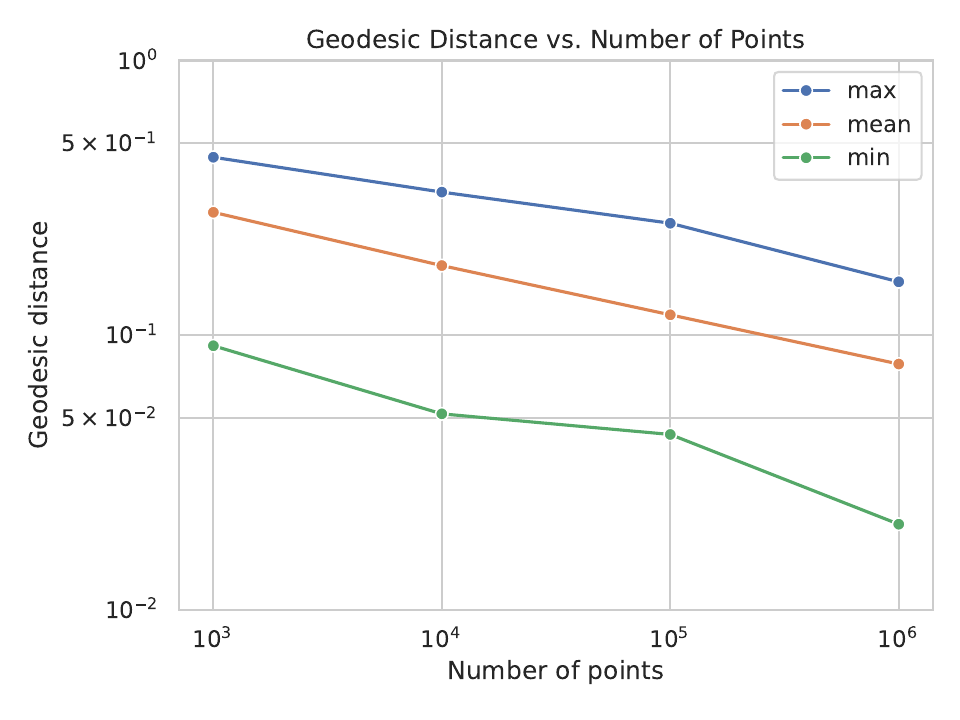}
\caption{\emph{Left:} Factor gained in the $\sigma$-loss as a function of average shortest geodesic distance between noised and sampled points. \emph{Right:} Average, maximum, and minimum geodesic distance as a function of the number of sampled points.}
\label{fig:expt}
\end{figure}

To do so, we need to compute the drop in the loss as a function of the geodesic distance between the test and train points. In practice, it requires sophisticated sampling techniques to produce a sample of test points at some fixed geodesic distance from the training set. For our purposes, instead of implementing such sampling, we approximate the same effect by adding Gaussian noise with varying variance to each training point. Hence the test points will only be ``approximately'' on the CY. The analysis effectively allows us to interpolate between testing on train points and testing on test points as a function of the variance of the Gaussian noise. The factor gained in the $\sigma$-loss versus average shortest geodesic distance between noised and sampled points is presented in the left plot of Figure ~\ref{fig:expt}; in the right plot, we show how the maximum, mean, and minimum geodesic distance changes with the total number of sampled points. The experiments were run with hyperparameters given in Table~\ref{table:hyper}.

\begin{table}[t]
    \centering
    \begin{tabular}{|c|c|c|c|}
    \hline
     & \textbf{ReLU} & \textbf{Gaussian} & \textbf{Delta} \\
    \hline
    \# Points & 2500 & 2500 & 1000 \\
    \hline
    Learning Rate & 0.1 & 0.1583 & .3162 \\
    \hline
    \texttt{$\sigma$} & - & 30.0 & - \\
    \hline
    $\sigma_b$  & 80.52 & - & - \\
    \hline
    $\sigma_w$  & 5.716 & - & - \\
    \hline
    \end{tabular}
    \caption{Hyperparameters for CY metric learning with various kernels. An empty entry indicates that the hyperparameter is not applicable to a given kernel.}
    \label{table:hyper}
\end{table}

From Figure \ref{fig:expt} we see a number of results. First, as the geodesic distance approaches zero, i.e., when using the train set as the test set, we see that the Gaussian and Delta kernels achieve a factor of $\sim 10^6$ in the $\sigma$-loss, even though there are only $\sim 1000$ train points; the ReLU kernel achieves a factor of $\sim 5$. However, as noise is added the test point become further away from the train points, corresponding to a growing geodesic distance. As the geodesic distance increases, the gain factor falls off, until the gain factor is only $\sim 1$ for distance $1$. This indicates that when the test points are too far from the train points, the kernel methods fail to generalize. From the right plot in Figure~\ref{fig:expt}, we see that the mean geodesic distance $\mu$ between points (and therefore train and test points, since they are sampled fairly from the Shiffman-Zelditch measure) is about $.3$ and $.1$ for $10^3$ and $10^5$ points, respectively. Comparing these distances to the plot on the left in Figure~\ref{fig:expt}, we see that with geodesic distances in this range we expect the gain factor to be about $5$, regardless of whether you train at $10^3$ points or $10^5$ points. We confirmed this with direct experimentation: at $10^3$ and $10^5$ points, the best our experiments did on test loss was about a factor of $5$. 

The advantage of the noise variance analysis is that it helps us understand what gains we might hope for in the test loss as we scale up the number of points. From the right plot in Figure~\ref{fig:expt} we see a linear dependence in the mean geodesic distance on a log-log scale. From the best linear fit to the data we obtain
\begin{align}
\label{eq:MeanDistance}
    \mu = .968 \times N_{\text{pts}}^{-0.184}.
\end{align}
Note that on theoretical grounds, one could expect $\mu$ to be proportional to $(N_{\text{pts}})^{-1/d}$ for a point sample on a (real) $d$-dimensional manifold. In our case, this would lead to an exponent of $-0.167$, which is close to the fitted value. The prefactor is more complicated and determining it analytically would require solving a sphere-packing type problem on the CY with respect to the geodesic distance using the FS metric and the Shiffman-Zelditch distribution. We just note that the mean line segment length between points on a 6d unit hypercube is $0.969$, which is numerically very close to the prefactor from our fit.

Extrapolating, to obtain the gain factor $\sim100$ achieved with finite neural networks requires a mean geodesic distance of $5 \times 10^{-3}$ for the Gaussian kernel, which in turn requires $10^{12}$ points. For the Delta kernel, a gain factor of $\sim 100$ requires a mean geodesic distance of $5\times 10^{-2}$, which in turn requires $10^{8}$ points. Since the finite neural network achieves this accuracy by learning from $10^5$ points, it is significantly more data efficient than the kernel methods. To obtain the gain factor of $\sim 10^6$ we observe for the training set also for the test points, requires $\sim 10^{35}$ points.

\vspace{1cm}
\noindent \emph{Discussion of the Importance of Feature Learning.}
The fact that the kernel methods fail to generalize well for larger distances is perhaps not too surprising. The NTK is fixed purely by the architecture and the parameter initialization. It is set in stone before training even starts and does not change ever. This means that, unlike a finite NN that can dynamically adapt to its inputs and learn useful embeddings for regression, the fixed kernel dictates how results at training points $x$ are combined to form a prediction at test points $x'$. 

The fixed kernel methods seem at odds with the fact that the Calabi-Yau metric is unique in a fixed K\"ahler class. If a kernel method is to be used to predict 
the metric at $x'$ given nearby metric information, it probably has to be a quite special kernel. Its analytic expression is likely a very complicated function and not a simple Gaussian or the NTK of a ReLU. In the only case where we know the CY metric, which is the trivial case of a two-torus, we can compute this function. For the analog of the Fermat Cubic, which is the 1d analog of the Fermat Quintic in~\eqref{eq:FermatQuintic}, this function is given in terms of inverse Weierstrass $\wp$-functions~\cite{Ahmed:2023cnw}, which is certainly not reproduced by adding metrics at nearby points weighted by a Gaussian or any natural guess for a kernel function. This makes the fact that the finite NN performs so well even more impressive.

\begin{figure}[t]
\centering
\includegraphics[width=.48 \textwidth]{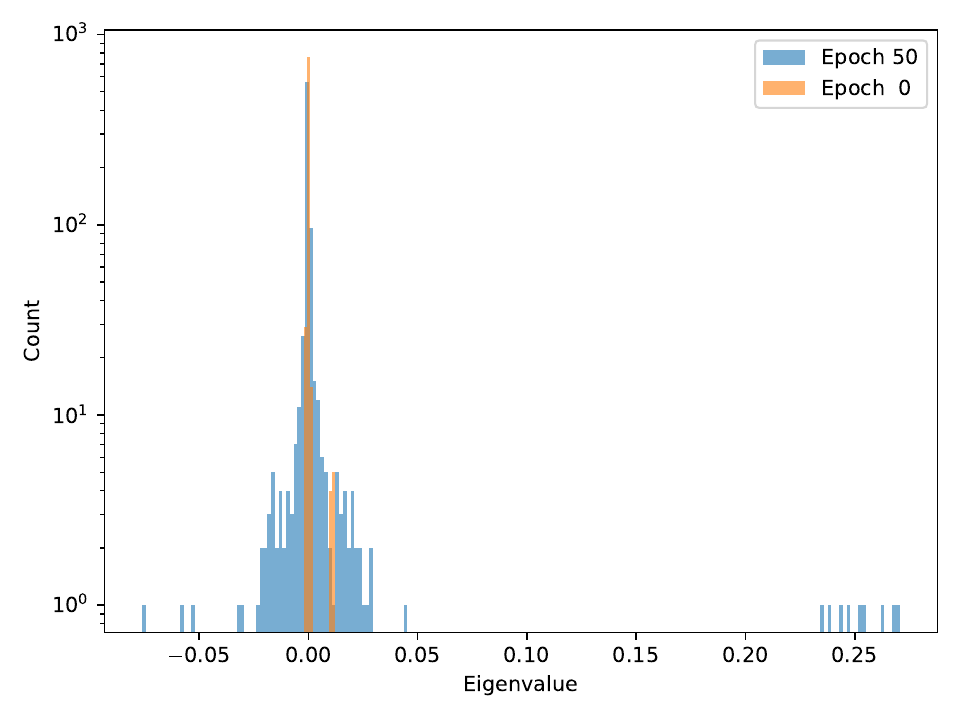} \quad
\includegraphics[width=.48 \textwidth]{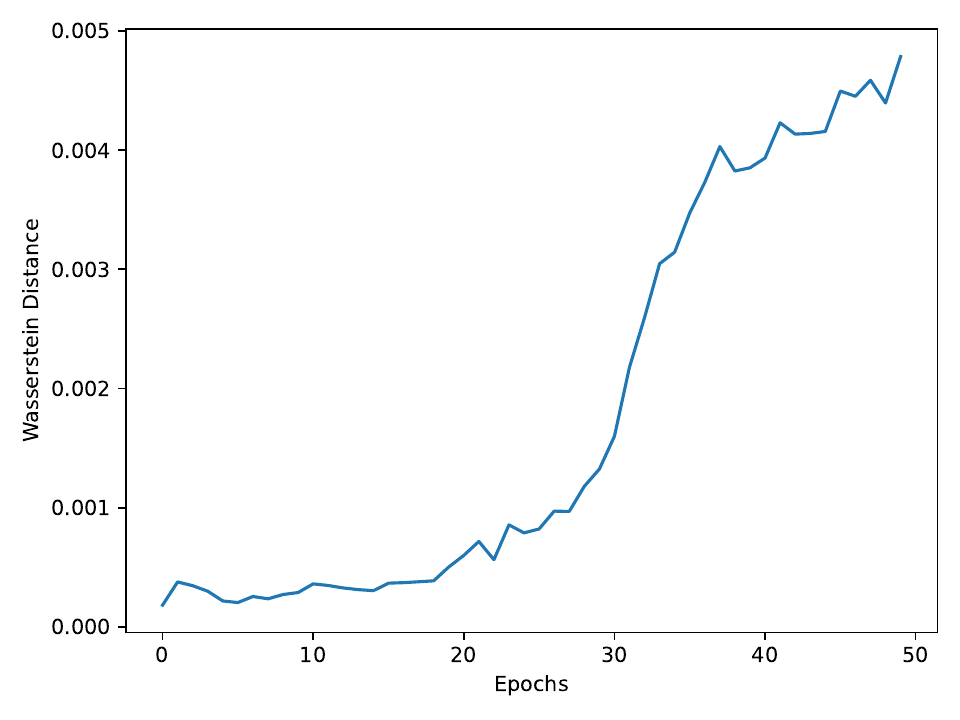}
\caption{\emph{Left:} Histogram of metric-NTK spectrum at Epochs $0$ and $50$, demonstrating clear distribution shift. \emph{Right:} Wasserstein distance between metric-NTK spectra at Epoch $0$ and stated epoch, demonstrating evolution during training.}
\label{fig:histograms}
\end{figure}

\subsection{Calabi-Yau Metric Learning and NTK Evolution}
\label{subsection:NTK_evolution}

Our numerical results demonstrate that the kernel learning associated to a frozen-NTK infinite width limit is not sufficient to learn the Calabi-Yau metric with a reasonable number of train points. This is in contrast to the finite width NNs of previous works, which can learn the metric efficiently, suggesting that feature learning is crucial for learning the metric. 

We can quantify this phenomenon by studying the evolution of the empirical metric-NTK during training: if the metric-NTK is frozen, features are not being learned. To quantify how much the NTK evolves during training, we study the so-called multiplicative model of the cymetric package~\cite{Larfors:2021pbb, Larfors:2022nep}, which approximates the metric components $g_{ij}$ using an MLP. We choose an architecture with three hidden layers with 128 neurons each and gelu activation function and train the model for 50 epochs. We compute the empirical NTK $\Theta_{ijkl}(x,x')$ of this model after each epoch for 10 randomly selected (but fixed) train set points $x$ and 10 randomly selected (but fixed) test set points $x'$. Since a hermitian $d\times d$ matrix has $d^2$ real degrees of freedom, we combine the indices $(ij)$ of the independent components into a multi-index $I$ and $(kl)$ into a multi-index $J$, with $1\leq I,J\leq9$. The NTK is thus a $9\times 9\times 10\times 10$ matrix. As a summary statistic, we compute the eigenvalues of the $10\times10$ matrix for each of the $9\times9$ entries $(I,J)$ and track their (combined over all $(I,J)$) evolution over the course of training. The results are presented in Figure \ref{fig:histograms}. We see that the eigenvalue spectrum of the metric-NTK evolves significantly during training, with a clear distribution shift between the spectra at epoch $0$ and epoch $50$. To quantify the shift in the distributions, we track the Wasserstein distance between the trained spectrum and the initial spectrum during training, showing significant evolution. Together, these plots demonstrate that the metric-NTK evolves significantly during training, and that the feature learning is crucial for the success of the NN in learning the metric.

\section{Conclusions}
We develop a theory for metric flows induced by neural network gradient descent. In general, the evolution of a NN that describes a metric is a complicated function of its hyperparameters, its parameters at initialization, and it evolves over the course of the training process. In the infinite width limit, this complicated training process becomes tractable and can even be described analytically in some cases. 

We illustrated how NNs can be engineered to reproduce flows with specific properties, such as Perelman's Ricci flow. The concept generalizes to other kinds of flows that are gradient flows of some energy functional, as reviewed in Appendix~\eqref{app:Flows}. This allows to engineer metric flows that are well-studied in the mathematics literature in terms of neural networks. Conversely, one can ask which type of flow a NN with a specific choice of architecture induces. Beyond that, one can study flows from kernel methods that are not the Neural Tangent Kernel of any known NN, leading to more general kernel methods. 

Numerically, we encounter multiple challenges. The most important one is that kernel methods require huge matrices that exceed the available computational resources of most users. To ameliorate this, we develop a batched feature clustering algorithm based on the geodesic distance in feature space. In general, there is no canonical metric for feature space, or even a natural choice of coordinates for the input data manifold. However, for the case at hand, the input data manifold is just the manifold for which we want to study the metric flow, such that there is a canonical choice of coordinates and reference metric in this coordinate system, with respect to which geodesic distances can be computed.

For the test case of the Fermat Quintic, we observe that the metric flow on the training set leads to almost Ricci-flat metrics. However, these results fail to generalize well for the kernels we tested. These included both Neural Tangent Kernels as well as other kernels that are not necessarily derived from a Neural Network. We attribute the fact that the results are worse than those obtained from a simple finite NN to the importance of feature learning. Kernel methods where the kernel is fixed during training cannot learn features. Conversely, we showed that finite neural networks that learn metrics well have metric-NTKs that evolve significantly in time.

The failures of infinite-NN kernel methods as compared to the performance of their finite counterparts is likely a consequence of the Calabi-Yau theorem: since the Calabi-Yau metric is unique, kernels that make correct predictions for the metric at $x'$ given nearby metric information must be very special. A priori such kernels would have nothing to do with the kernels associated to randomly initialized infinitely wide neural networks. Instead, for neural network kernel methods to work, the kernel should be learned so that it falls into this distinguished subclass. This is precisely what the finite neural networks achieve.

\bigskip
We end with some discussion of the main points. We have developed a theory of metric flows induced by neural network gradient descent, which is the main result. 

However,
the reader may find that the assumptions that lead to a realization of Perelman's Ricci flow --- infinite-width limit, locality, and elimination of component mixing --- are both ad hoc and strong. Indeed, this is the point: this rather famous metric flow arises as one instance in the much broader context afforded by neural network metric flows. Thus, the analytic tractability that was crucial for proving the Poincar\'e conjecture may correlate with specificity that is suboptimal for learning Calabi-Yau metrics efficiently. Conversely, the recent empirical successes of neural network approximations of Calabi-Yau metrics don't make \emph{any} of the assumptions that led to Perelman's Ricci flow (or other fixed kernel methods). The relative relationship between these good CY metric flows and Ricci flow is depicted in Figure \ref{fig:flow summary}, and begs the related question: are the infinite-width techniques also good for learning Calabi-Yau metrics?

Our results show that the answer is no, as might have also been expected on ML theory grounds (though there are some regimes in which kernel methods compete well with neural networks \cite{NEURIPS2020_ad086f59}). Specifically, we conduct experiments to approximate Calabi-Yau metrics with kernel methods of a similar type to those required for Ricci flow, and find that with a fixed number of points sampled on the Calabi-Yau, finite neural networks lead to far better performance. The key idea is that the fixed kernel do not learn features or evolve. A direct numerical analysis \cite{Jejjala:2020wcc} with Ricci flow also does not perform as well (beyond the torus) as finite neural network approaches. To make the converse point, we perform a finite neural network CY metric experiment that demonstrates the evolution of its metric-NTK.

In summary: we develop a general theory of metric flow induced by neural network gradient descent, and show the Ricci flow can arise in this context but only after making a number of very strong assumptions that lead to learning with a fixed kernel.  This suggests that the power of neural networks to learn Calabi-Yau metrics rests on their ability to learn features, explaining recent successes in the field.

\subsubsection*{Acknowledgements} 
We thank Robert Bryant, Michael Douglas, Mike Freedman, Sergei Gukov, Mark Haskins, Matt Headrick,  Jonathan Heckman, Cumrun Vafa, and Toby Wiseman for useful conversations. We are grateful to The Simons Center for Geometry and Physics for an enlightening program on computational differential geometry. J.H.\ is supported by NSF CAREER grant PHY-1848089. F.R.\ is supported by NSF grant PHY-2210333 and startup funding from Northeastern University. Both are supported by the National Science Foundation under Cooperative Agreement PHY-2019786 (The NSF AI Institute for Artificial Intelligence and Fundamental Interactions).

\appendix
\section{Primer on Neural Networks} 
\label{app:NNPrimer}
Deep neural networks (NNs) are the backbone of recent progress in machine learning and artificial intelligence, including image classification \cite{alexnet,inception} and generation \cite{dalle2}, gameplay such as in Chess \cite{AlphaZeroChess} and Go \cite{AlphaZero}, and natural language processing \cite{attention}.  

Neural networks, at their core, are functions. In most cases they are comprised out of sets of smaller functions, which are then composed with one another to form the so-called topology of the network, also known as the \emph{architecture}. This process typically involves the introduction of a set of parameters $\theta$ that are drawn from a distribution $P(\theta)$, written $\theta \sim P(\theta)$, at initialization, i.e., when the program begins. For fixed values of $\theta$ the network function
\begin{equation}
f_\theta: D \to R
\end{equation}
is a fixed function from some domain $D$ to some codomain $R$, typically $\bR^n$ and $\bR^m$; we will often suppress the subscript $\theta$, keeping in mind that neural networks always have parameters. 

Part of the progress in ML in the last decade is the development of architectures that are well-chosen for certain tasks.
A particularly famous architecture is the 
\emph{single-layer fully connected network}, also known as the \emph{perceptron}, which takes the form
\begin{equation}
f_i(x) = w^{(1)}_{ij} \sigma\left(w^{(0)}_{jk} x_k + b_j^{(0)}\right) + b_{i}^{(1)},
\end{equation}
where the matrices $w$ and vectors $b$ are parameters known as the weights and biases, superscripts denote the weights and biases of the $0^{\text{th}}$ and $1^\text{st}$ layer, and $\sigma:\bR \to \bR$ is an element-wise nonlinearity that is chosen as part of the architecture choice, e.g. $\sigma(z) = \tanh(z)$ or $\sigma(z) = \text{ReLU}(z):= \text{max}(0,z)$. Einstein summation is implied, and the full set of parameters is $\theta = \{w^{(1)}, b^{(1)}, w^{(0)}, b^{(0)}\}$. Our general sketch is that a neural network is a complicated function composed out of simpler functions according to the choice of architecture, and we see that sketch borne out here. Namely, we have
\begin{equation}
f: \bR^n \xrightarrow{w^{(0)}_{jk}(\cdot)+b^{(0)}_j(\cdot)} \bR^N \xrightarrow{\sigma(\cdot)} \bR^N \xrightarrow{w^{(1)}_{ij}(\cdot)+b^{(1)}_i(\cdot)} \bR^m,
\end{equation}
showing that the network is a parameter-dependent affine transformation composed with an elementwise non-linearity $\sigma$ and then another parameter-dependent affine transformation. The hyperparameter $N$ is known as the \emph{width} of the fully-connected network, but \emph{depth} can be added by sticking in more compositions of affine transformations and nonlinearities. More generally, a neural network is called a \emph{deep neural network} as the number of compositions is increased.

At initialization, a fixed network with fixed parameters $\theta$ is drawn from some distribution of functions. It is a random function that has not yet been tuned to any particular task. However, after initialization, the network is \emph{trained} to achieve some goal, such as those mentioned above, by updating the parameters $\theta$. The most famous update algorithm is \emph{gradient descent} (GD), which updates the parameters according to  
\begin{equation}
\frac{d\theta_I}{dt} =  - \frac{\partial {\cal L}[f]}{\partial{\theta_I}},
\end{equation}
where $\cL$ is a scalar loss functional that depends on $f$, and therefore on the parameters~$\theta$. We have written the continuum time equation, but in practice discrete time steps are taken on a computer. Typically $\cL$ is a sum of an element-wise loss $\ell$
\begin{equation}
\cL = \sum_{x' \in B} \ell[f](x'),
\end{equation}
where $B$ is a set of data points known as the \emph{batch}. A simple variant of gradient descent is \emph{stochastic gradient descent} (SGD), which at a fixed time step $t$ during training applies gradient descent to a random (hence, stochastic) proper subset $B_t \subset B$ known as the \emph{mini-batch}. While GD gets stuck when $\theta$ is at a critical point of the loss function, SGD often escapes the critical point due to the stochastic nature of the mini-batch.

Clearly this brief introduction only scratches the surface,  but it should aid in reading this paper. For the reader interested in a thorough introduction to neural networks, we recommend~\cite{Ruehle:2020jrk,bronstein2021geometric}.

\section{K\"ahler-Ricci flows for String Theory}
\label{app:Flows}
It has been recognized early on that a Ricci-flat manifold with external Minkowski space and a constant dilaton provide a solution to the supergravity equations of motion from string theory~\cite{Candelas:1985en}. Hence, Perelman's or Hamilton's Ricci flow can be used to find consistent background metrics for string compactifications. If there is no additional structure to the compactification space, and we are just interested in finding a Ricci-flat metric in a given topological class (e.g.\ for the case of M-theory compactifications on $G_2$ manifolds), this seems like a promising avenue.

In the case of compactifications on Calabi-Yau manifolds, we can make use of the fact that the metric is K\"ahler, which means it can be obtained from taking a holomorphic and an anti-holomorphic derivative of a real quantity $K$ called the K\"ahler potential,\footnote{The K\"ahler potential is only determined up to K\"ahler transformations, which means that $K$ is the section of a line bundle.}
\begin{align}
g_{a\bar b}=\partial_a\bar\partial_{\bar b} K\,,
\end{align}
subject to the constraint that the metric is positive definite. We can use this metric to define a closed (1,1)-form, the so-called K\"ahler form $J$, via
\begin{align}
J = \frac{i}{2}\, g_{a\bar b} \, dz^a \wedge d\bar z^{\bar b}
\end{align}
Due to this special structure, many quantities simplify. For example, the Christoffel symbols can only have 3 holomorphic or 3 anti-holomorphic indices. For the Ricci tensor, one finds the simple expression
\begin{align}
R_{a\bar b} = -\partial_a \bar \partial_{\bar b} \ln g\,
\end{align}
where $g = \det(g_{a\bar b})$. 
Inserting this in Hamilton's formulation of the Ricci-Flow, one finds
\begin{align}
\frac{dg_{a \bar b}}{dt} = \partial_a \bar \partial_{\bar b} \ln g\,.
\end{align}
We immediately see that the metric update for K\"ahler-Ricci flows is $\partial$-exact and $\bar \partial$-exact, and therefore the K\" ahler class is fixed under K\" ahler-Ricci flow.

We can further exploit the fact that the metric and all quantities that are derived from it are controlled by the $K$ and formulate the metric flow on the level of the K\"ahler potential, i.e., on the space\footnote{We want to point out two potential points of confusion that are owed to using standard symbols for quantities. The quantity $\varphi$ is not the same as the dilaton $\phi$. Likewise, the holomorphic top form $\Omega$ that will appear later is not related to the $\Omega$ appearing in the NTK kernel, and the loss function $\sigma$ (which is unrelated to a NN activation) is not related to any Gaussian width or noise variance.}
\begin{align}
\mathcal{H}=\{\varphi\in C^\infty(X)~|~J_0+\frac{i}{2\pi}\partial\bar\partial\varphi>0\}\,.
\end{align}
Here $J_0$ is a reference K\"ahler form (derived from a reference K\"ahler metric) which specifies the K\"ahler class of the metric we are interested in and is assumed to be constant throughout the flow. This means that the metric flow of
\begin{align}
g_{a\bar b}(t)=g_{0,a\bar b}+\frac{i}{2\pi}\partial\bar\partial\varphi(t)=\partial_a\bar\partial_{\bar b}(K_0+\frac{i}{2\pi}\varphi(t))
\end{align}
is induced by the flow of the K\"ahler potential correction $\varphi(t)$. This allows us to recast the problem of finding a Ricci-flat metric into the problem of solving a partial differential equation of Monge-Ampere type, which is what Yau used to prove Calabi's conjecture~\cite{Yau:1977aaa,Yau:1978aaa}. On the level of the flow, this leads to a parabolic flow equation in $\varphi$. 

\medskip
Just as in the case of the Perelman functional, whose functional variation gives rise to Hamilton flow (after a time-dependent diffeomorphism), there are several related functionals for flows in $\varphi$: the Mabuchi energy functional~\cite{Mabuchi:1986aa}, the  $J$-functional~\cite{Chen:2000aaa,Donaldson:2002aaa}, and the Calabi functional~\cite{Calabi:1982aaa}. The $J$-functional is the second term of the Mabuchi energy functional and the Calabi functional is the square of the derivative of the $J$-functional~\cite{Zheng:2015aaa}. We refer the reader to~\cite{Tian:2000ca,Song:2012aaa,Szekelyhidi:2014aaa} for more in-depth discussions of these quantities, and to~\cite{Headrick:2009jz} for a nice overview and application to CY metrics. 

The Calabi functional reads
\begin{align}
\label{eqn:CalabiFunctional}
E_\text{Calabi}=\int_X R^2 J^n\,,
\end{align}
where $R$ is the scalar curvature and $J^n$ is the integral measure. The flow associated with Calabi's functional~\eqref{eqn:CalabiFunctional} is
\begin{align}
\label{eqn:CalabiFlow}
\frac{d g_{a\bar b}}{d t}=\partial_a\bar\partial_{\bar b} R\,,\qquad\qquad R=g^{a\bar b} R_{a\bar b}=g^{a\bar b} \partial_a\bar\partial_{\bar b}\ln g\,.
\end{align} 
Note that the flow ends on metrics with constant curvature, $R=\text{const}$. For Calabi-Yau manifolds, this condition is equivalent to the seemingly stronger condition of Ricci flatness $R_{a\bar b}=0$~\cite{Song:2012aaa}:~a trivial anti-canoncial bundle means that the first Chern class of the Calabi-Yau manifold is zero, $c_1(X)=0$, which implies that the Ricci tensor is exact, $R_{a\bar b}=\frac{i}{2\pi}\partial_a\bar\partial_{\bar b} h$, for some function $h$. Taking the trace on both sides gives $\Delta h=0$, which means $h$ is constant and hence $R_{a\bar b}=0$.
Since $g_{a\bar b}=g_{0,a\bar b}+\frac{i}{2\pi}\partial_a\bar\partial_{\bar b}\varphi$, and $g_{0,a\bar b}$ is constant along the flow, the Calabi flow~\eqref{eqn:CalabiFlow} can be written (up to an integration constant which can be fixed from the initial conditions) as
\begin{align}
\frac{d \varphi}{d t} = g^{a\bar b} \partial_a\bar\partial_{\bar b}\ln g=\Delta \ln [\det(g_{0,c\bar d}+\partial_c\bar\partial_{\bar d}\varphi)]\,.
\end{align}

The $\varphi$-flows relevant for us has already been discussed in~\cite{Headrick:2009jz}, so we merely review it here following their exposition. As relevant background for utilizing their results, recall that a trivial anti-canonical bundle means that there exists a nowhere-vanishing holomorphic top form $\Omega$. In contrast to the Ricci-flat metric, analytic expressions for this $(n,0)$-form are known and can be computed straight-forwardly for Calabi-Yau manifolds that are given as complete intersections in toric ambient spaces~\cite{Witten:1985xc,Strominger:1985it}. From $\Omega$, one can construct a volume form $\mu_\text{CY}=(-i)^n\Omega\wedge\bar\Omega$. Since $h^{n,n}=1$ on a Calabi-Yau, this volume form is uniquely defined up to a constant. This means that $\mu_J=J^n/n!$, which is also an $(n,n)$-form, must be proportional to $|\Omega|^2$,
\begin{align}
J^n=\kappa|\Omega|^2\,.
\end{align}
In the literature, it is customary to define the sigma loss
\begin{align}
\label{eq:SigmaLoss}
\sigma :=|1-\eta|\,, \qquad\qquad\eta=\frac{J^n}{\kappa|\Omega|^2}\,.
\end{align}
Hence, the Calabi-Yau metric has $\eta=1$ and $\sigma=0$. 

The authors propose two classes of energy functionals: Ultra-local (meaning no derivatives) and two-derivative functionals. The ultra-local energy functionals are of the form
\begin{align}
E_\text{ultra}=\int_X F(\eta)\mu_\text{CY}\,,
\end{align}
for any differentiable, convex function $F:\mathbbm{R}^+\to\mathbbm{R}$ that is bounded from below. This has the variation
\begin{align}
\delta E_\text{ultra}=\frac12\int_X\mu_J\nabla^2F'(\eta) \varphi\,,
\end{align}
which implies that $F'(\eta)$ is constant. The authors propose using
\begin{align}
\label{eqn:HN}
F(\eta)=(\eta-1)^2\,,
\end{align}
since it is simple, contains no derivatives, and is convex. Choosing $F=\eta\ln\eta$, i.e., the (negative) von Neumann entropy of the probability distribution $\eta$ on $X$, leads to a gradient flow that is the Calabi flow~\eqref{eqn:CalabiFlow}.

For the two-derivative case, the authors consider
\begin{align}
\label{eqn:energy2Deriv}
E_\text{2deriv}=\int_X\mu_\text{CY}g^{a\bar b}\partial_a\eta\bar\partial_{\bar b}\eta\,,
\end{align}
which has the variation
\begin{align}
\delta E_\text{2deriv}=4\int_X\mu_\text{CY}\eta^{1/2}\partial^a\bar\partial^b\varphi\partial_a\bar\partial_{\bar b}\eta^{-1/2}\,.
\end{align}
As before, $E_\text{2deriv}$ is stationary iff $\eta$ is constant. Furthermore, since $J^n \propto \det(g_{a\bar b})$, we have $R_{a\bar b}=-\partial_a\bar\partial_{\bar b}\ln\eta$ and the functional~\eqref{eqn:energy2Deriv} becomes proportional to the Einstein-Hilbert action
\begin{align}
E_\text{EH}=-\frac12\int_X\mu_\text{CY} R\,.
\end{align}
This is also the variation of the $J$-functional and the ``square root'' of the Calabi functional. As explained before, the scalar curvature of a Calabi--Yau metric is zero and hence this functional is identically zero.

\bigskip

Let us finally come back to Ricci flow and the Perelman functional. Our goal is to realize Ricci flow as a $\varphi$-flow, and then realize $\varphi$-flow as a neural network metric flow. Setting $\phi=\ln\eta$ in~\eqref{eqn:PerelmanFunctional} (note that $\eta=1$ is constant for the Calabi-Yau, as is the dilaton $\phi$), the Perelman functional reads
\begin{align}
\mathcal{F}=-\int_X\mu_{\text{CY}} (R+g^{ab}\partial_a\ln\eta\; \partial_{b}\ln\eta)\,.
\end{align}
As we have just noted, $R+g^{a\bar b}\partial_a\ln\eta\;\bar\partial_{\bar b}\ln\eta=0$ for the Calabi-Yau case. This suggests that instead of the Ricci flow~\eqref{eqn:Ricci flow} obtained from the Perelman functional $\mathcal{F}$, one can take the flow corresponding to $E_\text{2deriv}$ in~\eqref{eqn:energy2Deriv}, which gives
\begin{align}
\frac{dg_{a\bar b}}{dt}=-\partial_a\bar\partial_{\bar b}\eta^{-1/2}\,,
\end{align}
or, using $g_{a\bar b}=g_{0,a\bar b}+\partial_a\bar\partial_{\bar b}\varphi$,
\begin{align}
\frac{d\varphi}{dt}=-\left(\frac{\det(g_{0,a\bar b}+\partial_a\bar\partial_{\bar b}\varphi)}{\mu_\text{CY}}\right)^{-1/2}\,,
\end{align}
Finally, the gradient flow associated with the ultra local energy functional~\eqref{eqn:HN} is
\begin{align}
\frac{d\varphi}{dt}=-2(\eta-1)=-2\left(\frac{\det(g_{0,a\bar b}+\partial_a\bar\partial_{\bar b}\varphi)}{\mu_\text{CY}}\right)\,.
\end{align}
This flow ends on the CY metric, for which $\eta=1$.

\bibliographystyle{bibstyle}
\bibliography{refs}

\end{document}